\documentclass{article}
\usepackage{amsmath}
\usepackage{amsfonts}
\usepackage{amssymb}
\usepackage{graphicx}

\input{tcilatex}

\begin{document}

\title{Global analysis of solar neutrinos ( assumed to be Majorana particles
) together with the new KamLAND data in the RSFP framework}
\author{D. Yilmaz and A.U. Yilmazer \\
Department of Physics Engineering,Ankara University\\
Tandogan,06100, Ankara TURKEY}
\maketitle

\begin{abstract}
Assuming neutrinos are Majorana particles having non-zero transitionmagnetic
moments, a global analysis of solar neutrino data, together with the new
KamLAND data, is presented in the resonant spin flavor precession (RSFP)
framework. We used Wood-Saxon and Gaussian shaped magnetic field profiles
throughout the entire Sun. For each magnetic field profile, allowed regions
from solar data combined with new KamLAND data are examined at 95\%
confidence level (CL). We showed that all allowed regions are in the large
mixing angle (LMA) region and remain the same as the $\mu B$ value is
increased for both magnetic field profiles, contrary to the Dirac case
studied in our previous work. The electron antineutrino flux from the Sun is
calculated, and via the results obtained several limits are set for $\mu B$.
\end{abstract}

\textbf{1.Introduction}\linebreak

The magnetic moment solution to solar neutrino problem is alternative
possible mechanism to MSW $\left[ 1-2\right] $ neutrino oscillation
solutions. In this solution motivated by Okun $et$ $al$ (OVV) $\left[ 3%
\right] $, strong magnetic field in the Sun changes neutrino's helicity
which has non-zero transition magnetic moment, as it passes through a region
with the magnetic field. Shortly after this solution was proposed, resonance
spin flavor precession (RSFP) was proposed by Lim and Marciano $\left[ 4%
\right] $. In the RSFP scenario, there is a combined effect of matter and
magnetic field in the Sun on neutrino's spin and flavor precession. Thus,
the neutrino's spin and flavor flip simultaneously. In other words,
left-handed electron neutrinos convert to right-handed neutrinos of other
types: $\nu _{e_{L}}\rightarrow $\ $\nu _{\mu _{R}}$ or $\nu _{\tau _{R}}$.
In the Majorana case $\nu _{\mu _{R}}$ is $\left( \nu _{\mu _{L}}\right) ^{C}
$ with C being the charge-conjugation and called $\overline{\nu }_{\mu }$.
Matter-enhanced spin-flavor precession of solar neutrinos with transition
magnetic moments for chlorine and gallium experiments was investigated in
detail by Balantekin $et$ $al$ $\left[ 5\right] $ for Majorana and Dirac
case. Raghavan $et$ $al$ investigated solar antineutrinos and gave the
theoretical framework for the solar antineutrino flux calculation $\left[ 6%
\right] $. In literature, one can find other studies about the solar
antineutrino and the experimental limit on its flux $\left[ 7-10\right] $.
In $[7]$, observation of the antineutrinos originated from $^{8}B$ neutrinos
in the Sun discussed and upper limits on the antineutrino flux are obtained.
They used the angular distribution of positrons emitted in the reaction of
the inverse beta decay. Obtaining positrons is a signal for the electron
antineutrinos. In that study, to find the upper limit, statistics of Super
Kamiokande (SK) and directionality of positrons from inverse beta decay were
taken into account. In $[8]$, another bound on the antineutrino flux was
obtained using SK\ solar data. The antineutrinos contribution to the SK
background and angular variations of them lead us to set an upper bound on
the antineutrino flux. To get a more accurate limit, positron angular
distribution and antineutrino asimetry was investigated. Gando $et$ $al$ $[9]
$ gave experimental results for the antineutrino flux. Their limits is $0.8\%
$ $(90\%CL)$ of the Standart Solar Models' neutrino flux. Miranda $et$ $al$ $%
[10]$ discussed spin flavor precession effect on solar neutrinos assumed to
be Majorana particle. From the KamLAND $[11]$ constraint on the solar
antineutrino flux, they put a limit on the Majorana neutrino transition
magnetic moment.

In this work, assuming the fact that all neutrinos are Majorana particles,
we present a global analysis of solar data combined with the new KamLAND
data $\left[ 12\right] $ in the RSFP scenario. We obtain allowed regions by
using standart least squares analysis on the oscillation parameter space, $%
\Delta m^{2}$ and $\tan ^{2}\theta $, space. Our results showed that all
allowed regions are in the large mixing angle (LMA) region and have the same
chi-square value as $\mu B$ value is increased for both magnetic field
profiles. After the combined analysis of solar and KamLAND\ data, we examine
the electron antineutrino flux at certain $\Delta m^{2}$ and $\tan
^{2}\theta $ values.

This paper is divided as fallows: in section 2 we give general information
on the evolution equation of Majorana neutrinos having transition magnetic
moment. In section 3 the relevant magnetic field profiles are given.
Detailed statistical analysis are given in section 4. In section 5 a
theoretical framework is given for the calculation of the solar antineutrino
flux. Finally, our results and conclusion are presented in section 6.

\textbf{2. Evolution equation for Majorana neutrinos in the RSFP framework }

For two generations, Majorana neutrino flavors are $\nu _{e}$ , $\nu _{\mu }$
, $\overline{\nu }_{e}$ , $\overline{\nu }_{\mu }$ . Altough Dirac neutrinos
have diagonal and off-diagonal magnetic moments, Majorana neutrinos have
only off-diagonal (transition) magnetic moments. For the Majorana neutrinos
that propagate through matter and in a transverse magnetic field $B$, the
evolution equation in the case of two generations is

\begin{equation}
i\frac{d}{dt}\left[ 
\begin{array}{c}
\nu _{e} \\ 
\nu _{\mu } \\ 
\overline{\nu }_{e} \\ 
\overline{\nu }_{\mu }%
\end{array}%
\right] =H\left[ 
\begin{array}{c}
\nu _{e} \\ 
\nu _{\mu } \\ 
\overline{\nu }_{e} \\ 
\overline{\nu }_{\mu }%
\end{array}%
\right] 
\end{equation}%
where $H$ is given by%
\begin{equation}
H=\left[ 
\begin{array}{cccc}
V_{e} & \frac{\Delta m^{2}}{4E}\sin 2\theta  & 0 & \mu ^{\ast }B \\ 
\frac{\Delta m^{2}}{4E}\sin 2\theta  & \frac{\Delta m^{2}}{2E}\cos 2\theta
+V_{\mu } & -\mu ^{\ast }B & 0 \\ 
0 & -\mu B & -V_{e} & \frac{\Delta m^{2}}{4E}\sin 2\theta  \\ 
\mu B & 0 & \frac{\Delta m^{2}}{4E}\sin 2\theta  & \frac{\Delta m^{2}}{2E}%
\cos 2\theta -V_{\mu }%
\end{array}%
\right] 
\end{equation}%
The matter potentials for a neutral unpolarized medium are given as 
\begin{equation}
V_{e}(t)=\frac{G_{f}}{\sqrt{2}}(2N_{e}-N_{n})\text{ \ \ \ \ \ \ }V_{\mu }=-%
\frac{G_{f}}{\sqrt{2}}N_{n}
\end{equation}%
where $N_{e}$ and $N_{n}$ are electron and neutron number densities in the
Sun, respectively, and the $G_{f}=1.16636\times 10^{-5}GeV^{-2}$. In
addition in the Sun electron and neutron number densities are well
approximated by $N_{e}\simeq 6N_{n}\simeq 2.4\times 10^{26}\exp
(-r/0.09R_{\odot })cm^{-3}$ $\left[ 13\right] $.

One can find $\nu _{e}\rightarrow \overline{\nu }_{\mu }$ resonance from the
Hamiltonian through the evolution equation. The spin-flavor conversion
submatrix for this transition is given by

\begin{equation}
\left( 
\begin{array}{cc}
V_{e} & \mu^{\ast}B \\ 
\mu B & \frac{\Delta m^{2}}{2E}\cos2\theta-V_{\mu}%
\end{array}
\right)
\end{equation}
Thus $\nu_{e}\rightarrow\overline{\nu}_{\mu}$ resonance condition is

\begin{equation}
N_{e}=\frac{6\sqrt{2}}{10G_{f}}\frac{\Delta m^{2}}{2E}\cos 2\theta 
\end{equation}%
using the aproximation $N_{e}\simeq 6N_{n}$. This value for the electron
density required for $\nu _{e}\rightarrow \overline{\nu }_{\mu }$ resonance
of Majorana neutrinos is greater than the corresponding value in the Dirac
case. The concequences of this difference will be seen in the allowed
regions behaviour in the neutrino oscillation parameter space. In our
analysis, we find results numerically by diagonalizing the Hamiltonian in
equation (2). The relevant method was discussed in $[5]$.

\begin{figure}[th]
\centering \includegraphics[width=4in]{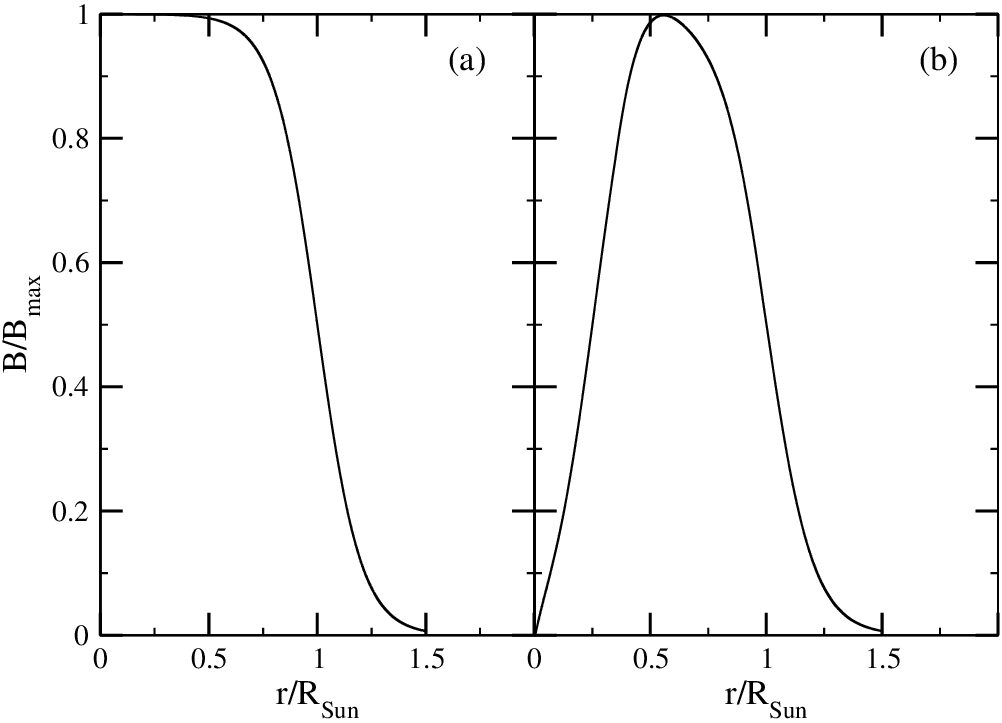}
\caption{Magnetic field profiles: (a) Wood-Saxon shape; (b) Gaussian shape.}
\label{fig:figure1.eps}
\end{figure}

\textbf{3. Magnetic field profile}

In the literature, various magnetic field profiles have been examined from
different aspects$\left[ 14-18\right] $. Here we considered only two as
typical for our analysis (see figure 1).

First, we took the magnetic field profile to be a Wood-Saxon shape of the
form

\begin{equation}
B(r)=\frac{B_{0}}{1+\exp[10(r-R_{\odot})/R_{\odot}]}  \label{3}
\end{equation}
where $B_{0}$ is the strength of the magnetic field at the center of the Sun.

The next magnetic field profile used in the analysis is of Gaussian
shape.\nolinebreak 

\textbf{4. Statistical analysis}

For completeness, we repeat the basics of the statistical methods of our
previous work. In this case, however, the relevant processes are to be made
carefully. In the literature, there is a common method often called $\chi
^{2}$ analysis to find the values of the neutrino oscillation parameters $%
\Delta m^{2}$, $\tan ^{2}\theta $ and to calculate the confidence levels of
allowed regions and the goodness of a fit $\left[ 19-22\right] $. In our
analysis, we use "covariance approach" to find the allowed regions mentioned
above. By this method, one minimizes the least-squares function

\begin{equation}
\chi _{_{\odot }}^{2}=\sum_{i_{1},i_{2}}^{N_{\exp }}(R_{i_{1}}^{(\exp
)}-R_{i_{1}}^{(thr)})(V^{-1})_{i_{1}i_{2}}(R_{i_{2}}^{(\exp
)}-R_{i_{2}}^{(thr)})  \label{4}
\end{equation}%
where $V^{-1}$ is the inverse of the covariance matrix of experimental and
theoretical uncertainties, $R_{i}^{(\exp )}$ is the event rate calculated in
the $i$th experiment and $R_{i}^{(thr)}$ is the theoretical event rate for
the $i$th experiment. The indices indicate the solar neutrino experiments: $%
i,i_{1},i_{2}=1,...,N_{\exp }$ with $N_{\exp }=4$.

Details of the expressions for theoretical event rates for all solar
neutrino experiments, chlorine experiments(Homestake) $\left[ 23\right] $,
gallium experiments (SAGE, GALLEX, GNO) $\left[ 24-26\right] $, Super
Kamiokande $[27]$ and SNO $[28,29]$ are given in $[30]$.

\begin{figure}[th]
\centering \includegraphics[width=4in]{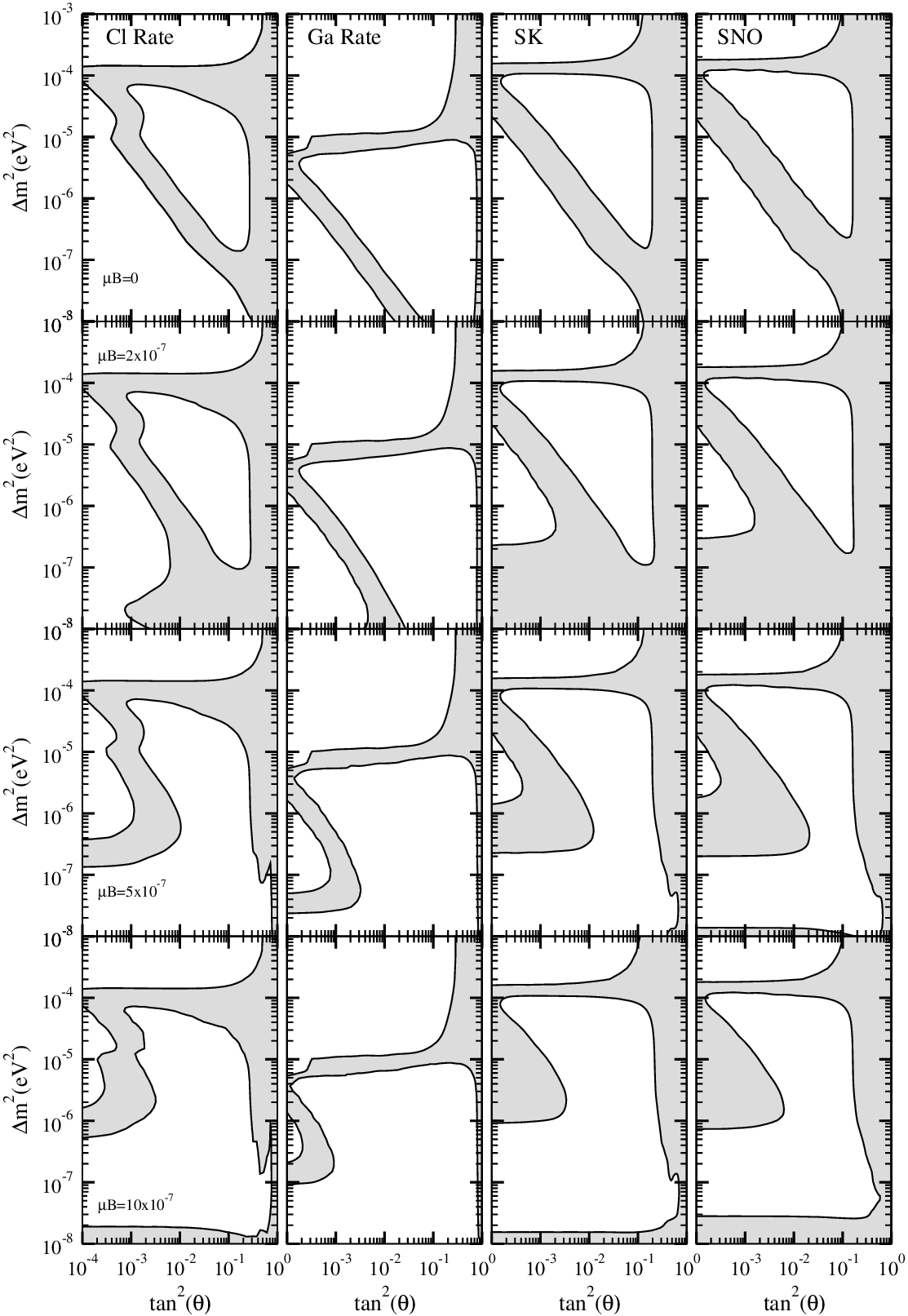}
\caption{The allowed regions for the Majorana case of neutrino parameter
space for each solar neutrino experiment seperately at $\protect\mu %
B=0,2,5,10\times 10^{-7}\protect\mu _{B}G$ and at 95\% CL. Each column and
row are for the same experiment and at the same $\protect\mu B$
value,respectively(e.g. in the second row at third column, an allowed region
for SK experiment at $\protect\mu B=2\times 10^{-7}\protect\mu _{B}G$ is
seen)}
\label{fig:figure2.eps}
\end{figure}

For the global analysis, we need $\chi _{KamLAND}^{2}$ $[12,31,32]$

\begin{equation}
\chi _{gl}^{2}=\chi _{_{\odot }}^{2}+\chi _{_{KamLAND}}^{2}
\end{equation}

We took fluxes and cross sections for event rates and error matrices from
Bahcall $[33]$.

\begin{figure}[ht]
\centering \includegraphics[width=4in]{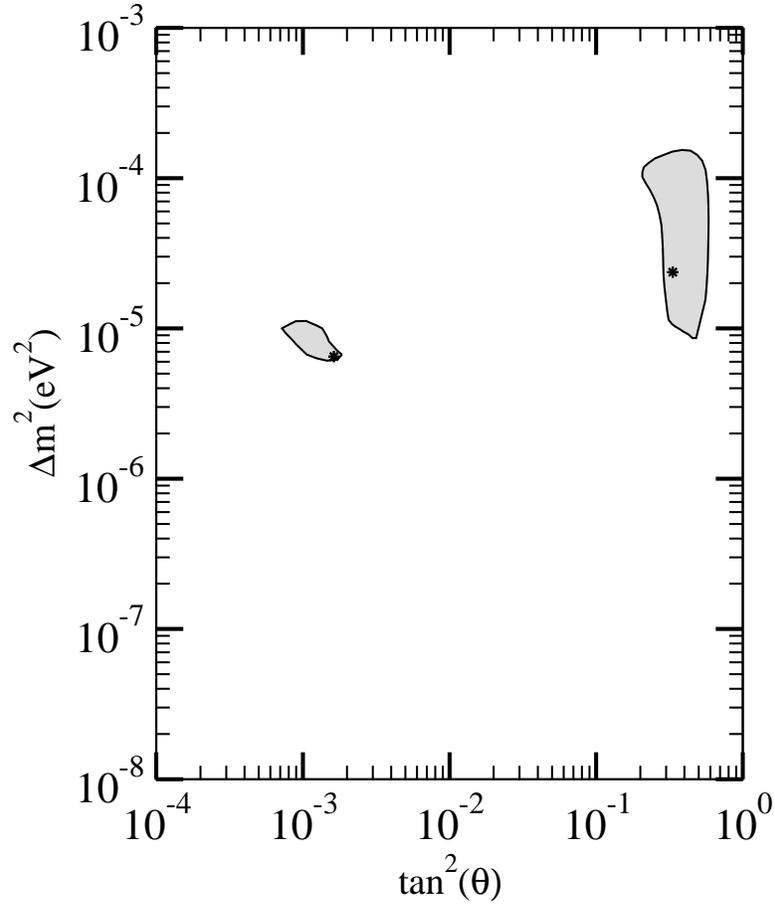}
\caption{The allowed regions from combined solar neutrino experiments(
chlorine, all three gallium, SK\ and SNO experiments) at the same $\protect%
\mu B$ values and CL as in figure 2. Stars indicate the local best-fit
points.}
\label{fig:figure3.eps}
\end{figure}

\begin{figure}[th]
\begin{center}
\includegraphics[width=4in]{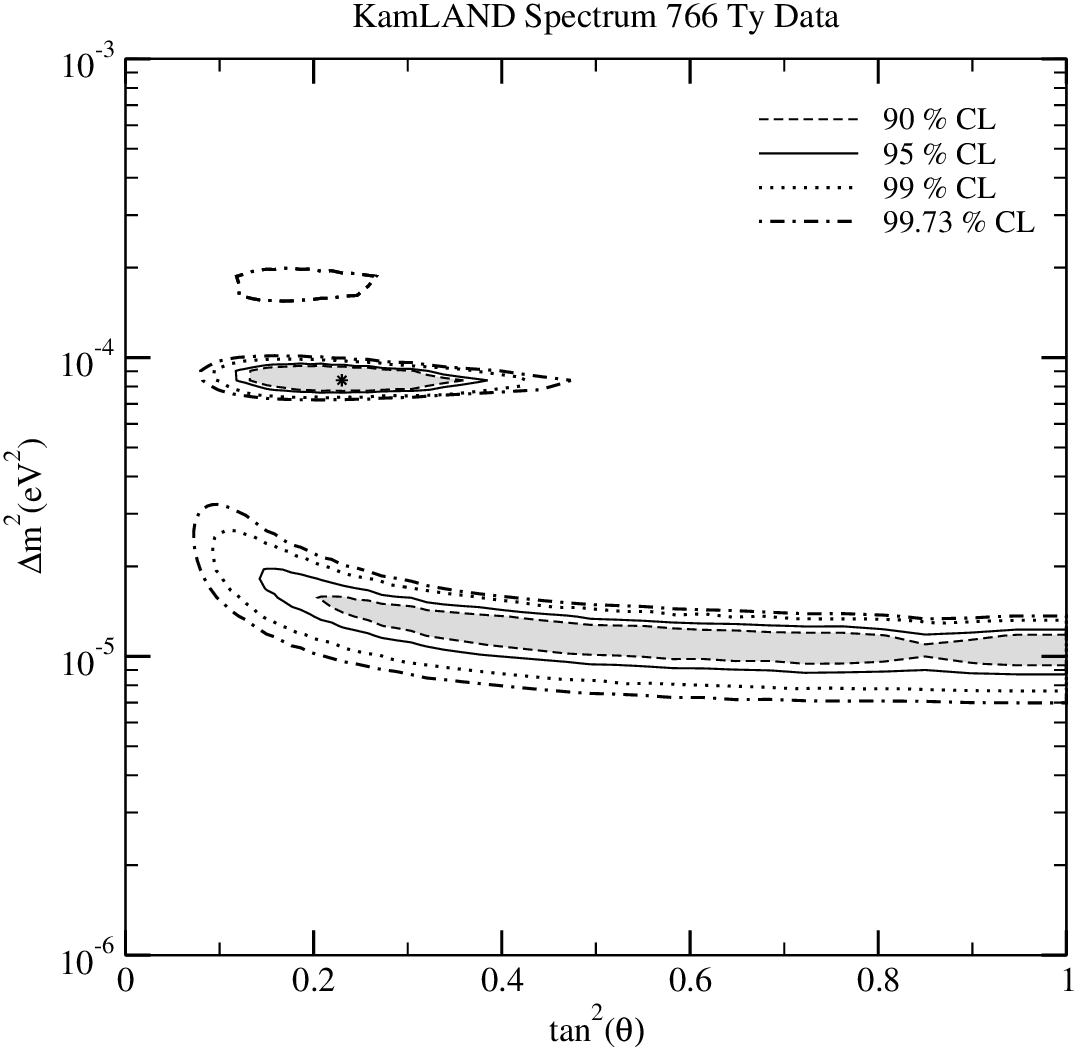}
\end{center}
\caption{Allowed region from the new KamLAND spectrum at different
confidence levels. Star indicates the best fit point.}
\label{fig:figure4.eps}
\end{figure}

\textbf{5.Solar electron antineutrino flux}

If neutrinos are Majorana type, $\nu _{e}$ changes to $\overline{\nu }_{\mu }
$ inside the Sun by SFP. After Sun, vacuum oscillation yields $\overline{\nu 
}_{\mu }\rightarrow $ $\overline{\nu }_{e}$. This process is given
schematically:

\begin{equation*}
\nu _{e}\overset{B_{Sun}}{\rightarrow }\overline{\nu }_{\mu }\overset{V_{osc}%
}{\rightarrow }\overline{\nu }_{e}
\end{equation*}

To find the electron antineutrino flux on Earth, one need to probability of $%
\nu _{e}\rightarrow \overline{\nu }_{e}$ transition, $P(\nu _{e}\rightarrow 
\overline{\nu }_{e})$:

\begin{equation*}
P(\nu _{e}\rightarrow \overline{\nu }_{e})=P(\nu _{e}\rightarrow \overline{%
\nu }_{\mu };SFP)\times P(\overline{\nu }_{\mu }\rightarrow \overline{\nu }%
_{e};VacuumOsc.)
\end{equation*}%
where $\nu _{e}\rightarrow \overline{\nu }_{\mu }$ transition probability $%
P(\nu _{e}\rightarrow \overline{\nu }_{\mu };SFP)$ is calculated numerically
from the equation (2). $P(\overline{\nu }_{\mu }\rightarrow \overline{\nu }%
_{e};VacuumOsc.)$ is the well known vacuum oscillation probability given as

\begin{equation*}
P(\overline{\nu }_{\mu }\rightarrow \overline{\nu }_{e};VacuumOsc.)=\sin
^{2}2\theta \sin ^{2}(\frac{\Delta m^{2}}{4E}R)\overset{averaging}{%
\rightarrow }\frac{1}{2}\sin ^{2}2\theta 
\end{equation*}%
where R is the distance between Sun and Earth.

Electron antineutrinos is detected observing the positron from the $%
\overline{\nu }_{e}+p\rightarrow e^{+}+n$ reaction. Positron event rate is
found from

\begin{equation*}
N=Q_{0}\int dE_{V}\int \epsilon (E_{e})R(E_{V},E_{e})\Phi _{\overline{\nu }%
_{e}}(E)\sigma (E)dE
\end{equation*}%
where $Q_{0}$ is a normalization constant taking into account the number of
atoms in the fiducial volume of the detector and its live time exposure.
Positron energy $E_{e}$ is%
\begin{equation*}
E_{e}=E-1.293MeV
\end{equation*}%
and visible energy, $E_{V}=E_{e}+m_{e}$. $\epsilon (E_{e})$ and $%
R(E_{V},E_{e})$ are the detection efficiency and the energy resolution
function of detector, respectively. $\sigma (E)$ is the antineutrino cross
section and $\Phi _{\overline{\nu }_{e}}(E)$ is the electron antineutrino
flux given by

\begin{equation*}
\Phi _{\overline{\nu }_{e}}(E)=\Phi _{\nu _{e}}(^{8}B)\times P(\nu
_{e}\rightarrow \overline{\nu }_{e})
\end{equation*}

\textbf{6.Results and conclusions}

In our calculations, we assumed that the magnetic field extends over the
entire Sun for both magnetic field profiles. All of the calculations have
been performed for both of the magnetic field profiles. Neutrino spectra are
taken from the Standard Solar Model of Bahcall and his collaborators $\left[
33\right] $. All allowed regions were calculated at 95\% CL in this work.

\begin{figure}[th]
\centering \includegraphics[width=4in]{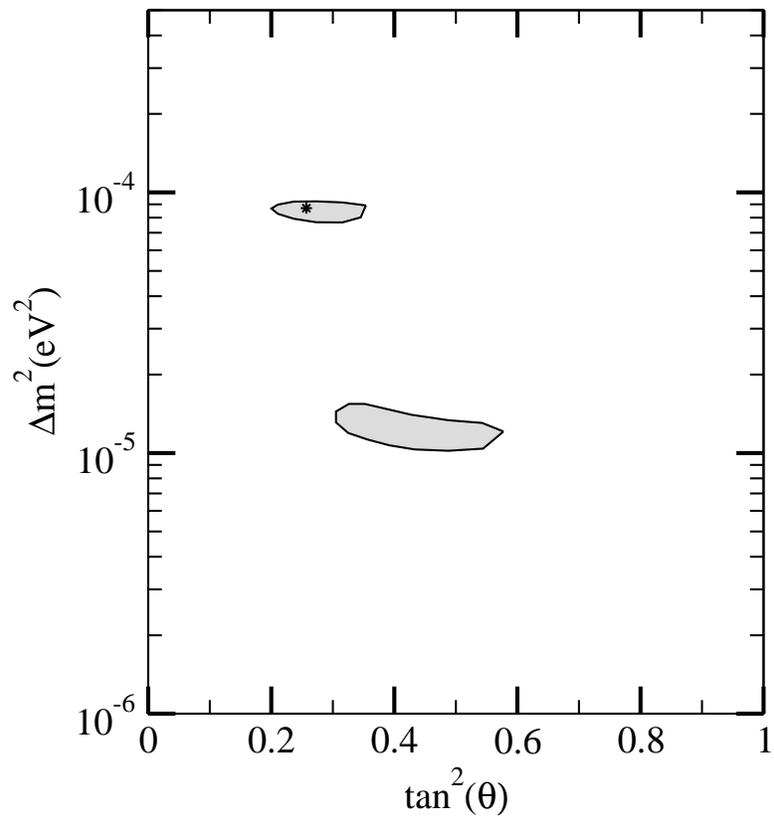}
\caption{Allowed regions from the combined solar plus new KamLAND spectrum
analysis at different $\protect\mu B$ values ($\protect\mu B=0,2,5,10\times
10^{-7}\protect\mu _{B}G$)and at 95\% CL. Regions and the best fit point
remined the same for all $\protect\mu B$ values. The star indicates the best
fit point.}
\label{fig:figure5.eps}
\end{figure}

First, we considered only solar neutrino data assuming neutrinos are
Majorana type and examined allowed regions for all solar neutrino
experiments ( Homestake, Gallium, Super-Kamiokande (SK) and SNO ) using
covariance approach of statistical analysis. These results are shown in
Figure 2 in which each column and row are for the same experiment and at the
same $\mu B$ value respectively(e.g. in the second row at third column, an
allowed region for SK experiment at $\mu B=2\times10^{-7}\mu_{B}G$ is seen).

We displayed the allowed regions from combined solar neutrino experiments in
Figure 3. In that figure for both magnetic field profiles, two local best
fit points are shown for $\mu B$ values that are $0,2,5,10\times 10^{-7}\mu
_{B}G$. The best fit points are at ( \ $tan^{2}\theta =1.63\times 10^{-3}$, $%
\Delta m^{2}=6.49\times 10^{-6}$eV$^{2}$ ) for $\mu B$ = $0,2,5\times
10^{-7}\mu _{B}G$ and ( \ $tan^{2}\theta =1.41\times 10^{-3}$, $\Delta
m^{2}=6.49\times 10^{-6}$eV$^{2}$ ) for $\mu B$ = $10\times 10^{-7}\mu _{B}G$%
, respectively.

After this step, we examined the new KamLAND data and displayed allowed
regions from new binned KamLAND data $[12]$ in Figure 4.

As a next stage, our global analysis combining solar and KamLAND spectrum
analysis were shown in Figure 5 for the same $\mu B$ values in Figure 2.
Regions and the best fit point remained the same for those $\mu B$ values,
such that, the best fit point is at \ $tan^{2}\theta =0.26$ and at $\Delta
m^{2}=8.66\times 10^{-5}$eV$^{2}$. All minimum chi-squares also remained the
same for those $\mu B$ values.

It was observed that there are no appreciable differences between the
results for the two magnetic field profiles we used; we have given only the
figures for Gaussian case.

In contrast to Dirac case $[34]$ in the Majorana case we investigated in
this paper, it seems that there is no appreciable magnetic field effect on
the allowed regions for Majorana neutrinos in our calculations.

\begin{figure}[th]
\centering \includegraphics[width=4in]{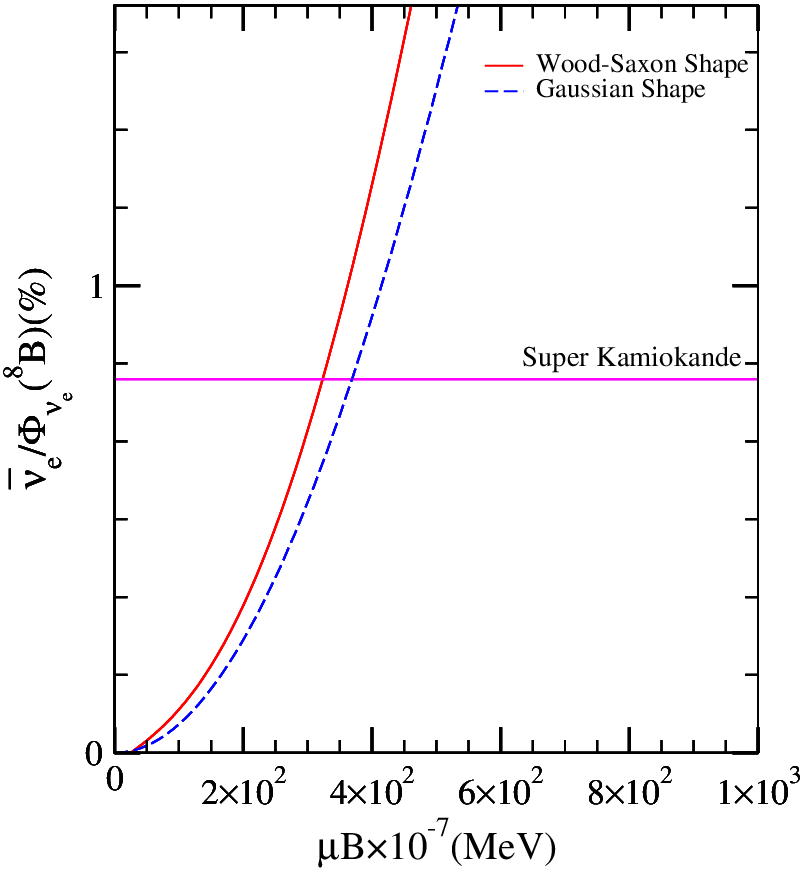}
\caption{Expected $\overline{\protect\nu }_{e}/\Phi _{\protect\nu %
_{e}}(^{8}B)$ for $\protect\mu B$ values. The horizantal line shows the
upper $\overline{\protect\nu }_{e}/\Phi _{\protect\nu _{e}}(^{8}B)$ flux
limit at SK.}
\label{fig:figure6.eps}
\end{figure}

Finally to put an upper limit on $\mu B$ for Majorana neutrinos, finally, we
calculated resulting electron antineutrino flux from the Sun and compared
with the limit obtained by Super Kamiokande. Our results are given in $%
\overline{\nu }_{e}/\Phi _{\nu _{e}}(^{8}B)-\mu B$ plane for 8-20 MeV
visible energy region and at best fit parameters $tan^{2}\theta =0.41$, $%
\Delta m^{2}=7.1\times 10^{-5}$eV$^{2}[35]$ in Figure 6. In that figure, the
horizantal line shows the upper $\overline{\nu }_{e}/\Phi _{\nu _{e}}(^{8}B)$
flux limit at SK $[9]$. From that limit, one can put a bound on $\mu B$ at
these values of $\Delta m^{2}$ and $tan^{2}\theta $. From our results, $\mu
B<3.2\times 10^{-5}\mu _{B}G$ and $\mu B<3.7\times 10^{-5}\mu _{B}G$ for
Wood-Saxon and Gaussian shaped magnetic field profiles, respectively.

\bigskip

\textbf{Acknowledgments}

We would like to thank to Prof.Dr. A. Baha Balantekin for his suggestions
and for his kind help.

\bigskip

\textbf{References}

$\left[ 1\right] $ Wolfenstein L, \textit{Phys. Rev. D} \textbf{17}, 2369
(1978) ; \textbf{20}, 2634 (1979).

$\left[ 2\right] $ Mikheyev S P and Smirnov A Yu , \textit{Nuovo Cimento} C 
\textbf{9}, 17 (1986) ; \textit{Yad. Fiz.} \textbf{42}, 1441 (1985) ; [Sov. 
\textit{J. Nucl. Phys.} \textbf{42}, 913 (1985)].

$\left[ 3\right] $ Okun L B, Voloshin M B, and Vysotsky M I,\textit{\ Yad.
Fiz. }\textbf{44}, 677 (1986) ; [\textit{Sov. J. Nucl. Phys}. \textbf{44},
440 (1986)].

$\left[ 4\right] $ Lim C S and Marciano W J, \textit{Phys. Rev.} D \textbf{37%
}, 1368 (1988)

$\left[ 5\right] $ Balantekin A B, Hatchell P J, Loreti F 1990 \textit{Phys.
Rev.} D \textbf{41} 3583.

$\left[ 6\right] $ Raghavan \textit{et al} 1991, \textit{Phys. Rev.} D 
\textbf{44} 3786.

$\left[ 7\right] $\ Fiorentini G, Moretti M, Villante F L 1997 \textit{%
Phys.Lett. B }\textbf{413} 378 (\textit{Preprint }astro-ph/9707097)

$\left[ 8\right] $\ Torrente-Lujan 2000 \textit{Phys.Lett. B }\textbf{494}
255 (\textit{Preprint }hep-ph/9911458)

$\left[ 9\right] $\ Gando Y \textit{et al} 2003 \textit{Phys.Rev.Lett.} 
\textbf{90} 171302 (\textit{Preprint }hep-ex/0212067)

$\left[ 10\right] $\ Miranda O G, Rashba T I,Rez A I, Valle J W F 2004 
\textit{Phys.Rev.Lett.} \textbf{93 }051304 (\textit{Preprint }hep-ph/0311014)

$\left[ 11\right] $\ (KamLAND Collaboration), Proposal for US participation
in KamLAND, 1999

$\left[ 12\right] $ \ \ Araki T \textit{et al} (KamLAND Collaboration) 2005 
\textit{Phys. Rev. Lett.} \textbf{94} 081801\textit{(Preprint hep-ex/0406035)%
}

$\left[ 13\right] $ Bahcall J N 1989\textit{\ Neutrino Astrophysics}
(Cambridge: Cambridge University Press)

$\left[ 14\right] $ Pulido J, 2002 \textit{A. High Energy Phys. }AHEP2003/046

$\left[ 15\right] $ Chauhan C B, Pulido J 2002 \textit{Phys.Rev.D }\textbf{66%
} 053006 (\textit{Preprint} hep-ph/0206193)

$\left[ 16\right] $ Chauhan C B, 2002 \textit{Preprint} hep-ph/0204160

$\left[ 17\right] $ Bykov A A, Popov V Y, Rashba T I, Semikoz V B, 1999 
\textit{Preprint} hep-ph/0002174,

\ \ \ \ \ \ Talk given at 10th International School on Particles and
Cosmology,

\ \ \ \ \ \ Karbardino-Balkaria, Russia, 19-25 Apr 1999 and at International
Workshop

\ \ \ \ \ \ on Strong Magnetic Fields in Neutrino Astrophysics, Yaroslavl,
Russia, 5-8 Oct 1999.

$\left[ 18\right] $ Akhmedov E Kh, Pulido J, \textit{Astropart.Phys. }%
\textbf{13} 227 (\textit{Preprint} hep-ph/9907399)

$\left[ 19\right] $ Feldman G J and Cousins R D 1998 \textit{Phys. Rev.} D 
\textbf{57 }3873 \textit{(Preprint physics/9711021)}

$\left[ 20\right] $\ Fogli G L, Lisi E, Marrone A, Montanino D and Palazzo A
2002 \textit{Phys. Rev.} D\textbf{\ 66} 053010 \textit{(Preprint
hep-ph/0206162) }

$\left[ 21\right] $ Garzelli M V and Giunti C 2002 \textit{Astropart. Phys}. 
\textbf{17} 205 \textit{(Preprint hep-ph/0007155)}

\ \ \ \ \ \ \ Garzelli M V and Giunti C 2002 \textit{Phys. Rev. }D \textbf{65%
} 093005 \textit{(Preprint hep-ph/0111254)}

\ \ \ \ \ \ \ Garzelli M V and Giunti C 2001\textit{\ J. High Energy Phys. }%
JHEP12(2001)017 \textit{(Preprint hep-ph/0108191) }

$\left[ 22\right] $\ Gonzalez-Garcia M C and Nir Y 2003 \textit{Rev.Mod.Phys.%
} \textbf{75} 345 (\textit{Preprint }hep-ph/0202058)

$\left[ 23\right] $ \ \ Cleveland B T \textit{et al }1998 \textit{Astrophys.
J}. \textbf{496} 505

$\left[ 24\right] $ \ \ Abdurashitov J N \textit{et al} (SAGE Collaboration)
2002 \textit{J. Exp. Theor. Phys.}\textbf{\ 95} 181

\ \ \ \ \ \ \ Abdurashitov J N \textit{et al} 2002 \textit{Zh. Eksp. Teor.
Fiz. }\textbf{122} 211 \textit{(Preprint astro-ph/0204245)}

$\left[ 25\right] $ \ \ Hampel W \textit{et al }(GALLEX Collaboration) 1999 
\textit{Phys. Lett. B} \textbf{447} 127

$\left[ 26\right] $ \ \ Altmann M \textit{et al} (GNO Collaboration) 2000 
\textit{Phys. Lett. }B \textbf{490} 16 \textit{(Preprint hep-ex/0006034)}

$\left[ 27\right] $ \ \ Fukuda S \textit{et al }(Super-Kamiokande
Collaboration) 2001 \textit{Phys. Rev. Lett.} \textbf{86} 5651 \textit{%
(Preprint hep-ex/0103032) }

$\left[ 28\right] $ \ \ Ahmad Q R \textit{et al} (SNO Collaboration) 2001 
\textit{Phys. Rev. Lett}. \textbf{87 }071301 \textit{(Preprint
nucl-ex/0106015)}

$\left[ 29\right] $ \ \ Ahmad Q R \textit{et al }(SNO Collaboration) 2002 
\textit{Phys. Rev. Lett}. \textbf{89} 011301 \textit{(Preprint
nucl-ex/0204008) }

$\left[ 30\right] $ \ \ Balantekin A B, Yuksel H 2003 \textit{J. Phys.} G 
\textbf{29}, 665 

$\left[ 31\right] $ Murayama H and Pierce A, \textit{Phys. Rev.} D \textbf{65%
} 013012 \textit{(Preprint hep-ph/0012075) }

$\left[ 32\right] $ Bandyopadhyay A, Choubey S, Goswami S, Gandhi R, Roy D
P, \textit{J. Phys.} G \textbf{29}, 665 (2003)

$\left[ 33\right] $ \ \ Bahcall J N, Pinsonneault M H and Basu S 2001 
\textit{Astrophys. J.} \textbf{555} 990 \textit{(Preprint astro-ph/0010346). 
}

\ \ \ \ \ \ \ See also Bahcall's homepage,\ http://www.sns.ias.edu/\symbol{%
126}jnb/ \textit{\ }

$\left[ 34\right] $ \ Yilmaz D, Yilmazer A 2005 \textit{J. Phys.} G \textbf{%
31}, 57 

$\left[ 35\right] $ Ahmed S N\ \textit{et al} 2004 \textit{Phys. Rev. Lett}. 
\textbf{92 }181301

\end{document}